\documentclass[manuscript]{aastex}
\slugcomment{ApJ, accepted (6/27/2012)}
\usepackage{natbib}
\shortauthors{Focardi \& Malavasi}
\shorttitle{Faber Jackson and environment}

\begin{document}

\title{The effect of the environment on the Faber Jackson relation}

\author{ P. Focardi\altaffilmark{1} and N. Malavasi\altaffilmark{1}}
\affil{Dipartimento di Astronomia, Universit\'a di Bologna, Italy}

\begin{abstract}
We investigate the effect of the environment on the Faber Jackson (FJ) relation,
using a sample of 384 nearby elliptical galaxies and estimating objectively 
their environment  on the typical scale of galaxy  clusters. 
We show that the intrinsic scatter of the FJ 
is significantly reduced when ellipticals in high density environments are
compared to ellipticals in low density ones.
This result, which holds on  a limited range of overdensities, is likely to provide an 
important observational 
  link between scaling relations and formation mechanisms in galaxies.
   
\end{abstract}


\section{Introduction}

The Faber Jackson (FJ)  is the first scaling relation discovered
for elliptical galaxies. Already described by Morgan \& Mayall (1957), although in a qualitative way,
 it was given its first quantified form 20 years later
 by Faber \& Jackson (1976) who, on the basis of a handful of nearby 
early type galaxies, were able to proof the existence of a power law relation linking
luminosity ($L_{\rm B}$) to central velocity dispersion ($\sigma_0$). 
Soon thereafter Kormendy (1977) found a second scaling relation which holds for elliptical galaxies, 
relating the effective surface brightness ($\mu_{\rm e}$) to the effective radius ($R_{\rm e}$). The Kormendy relation
was refined  ten years later for ellipticals by Hamabe \& Kromendy (1987) and in that  same year Dressler et al (1987) and Djorgovsky \& Davis (1987) 
 discovered a more general relation (the Fundamental Plane, FP) 
linking  Log $R_{\rm e}$, Log $\sigma_0$ and $\mu_{\rm e}$. 

The scaling relations are powerful tools that can be used  to derive galaxy distances and, even more important, 
constitue an invaluable observational  bench mark for theoretical models. It is especially for this latter reason  that  they 
have been the subject of much interest since their discovery. Understanding origin and nature of the scaling relations is a fundamental quest for any successful
theory of galaxy formation which is expected to be able to predict the observed slope, scatter, possible variation (as a function of luminosity, wavelength, environment) and 
evolution (with z). The much narrower scatter displayed by the FP with respect to the FJ and Kormendy relations
made the former one to become rapidly more attractive than the latter two which were easily  interpreted  to be partial representations (projections) of the FP onto a 
lower dimensional space \cite{dr1987,dj1987,fa1987,dz1991}.

  Very recently the FJ appears to have captured again attention on the theoretical point of view, as Sanders (2010) has claimed it to be
  more fundamental and universal than the FP within the context of MOND (modified Newtonian dynamics). 
  This finding is expected to motivate renewed interest in the FJ, which so far has not been largely investigated. 
   There is no much  work which has been carried out  on 
  the FJ relation if one excepts studies which have provided evidence for a decrease of its stepness 
  at low luminosity \cite{to1981,da1983,he1992,fr2005,ma2005,be2006,ds2007,la2007,vo2007,ko2012} and studies devoted to investigate
  the effect of  luminosity, mass and redshift on it \cite{fr2005,be2006,ds2007,ni2010,ni2011}. 
  At variance with the FP for which the effect of the environment has been largely investigated,
  although  with rather conflicting results 
  \cite{de1992,ma1996,ma1999,de2001,tr2001,be2003,ev2002,go2003,re2004,re2005,de2005,do2008,la2010}, 
  so far only 4 studies exist \cite{zi2005,fr2005,fr2009,fz2009} which have looked for possible effects induced by the environment on the FJ relation of
  rather distant (z $\in $ [0.2 -- 0.7]) early-type galaxies, without finding however any strong evidence for them. 
  
  According to the standard cosmological paradigm, structures in the present day Universe have formed 
  through a hiearchical scenario process predicting rather different assembling time scales and evolutionary paths for galaxies in high
  and low density regions (Baugh et al. 1996; Kauffmann \& Charlot 1998; Somerville \& Primack 1999; Kauffmann et al. 2004).
  Environment is thus expected to play a relevant role in shaping galaxy properties and is likely to  leave its imprint in the scaling relations
  as well.  
  This is the reason which has motivated the above mentioned studies (mostly
  concentrated on the FP) and the present work devoted to investigate the effect of the environment on the FJ relation, using a sample of 384 nearby ellipticals and estimating
  their environment on the typical scale of galaxy clusters. 
  
  The
  structure of the paper is the following: in $\S$2   we present the sample, in section $\S$3 we derive the FJ relation for the whole sample and for its bright and faint
  components and test the robustness of our results accounting both for errors on $\sigma_0$ and $m_{\rm B}$, in $\S$4 we illustrate the method that we 
  have used to estimate the environment, in $\S$5 we show that the scatter of the FJ gets largely reduced in high density environments 
  and increased in low density ones 
  and that this difference is neither induced by errors on $\sigma_0$  nor by luminosity difference between the samples,
  in  $\S$6 we show that the scatter of the FJ relation increases with decreasing density in overdense environments and decreases with increasing density in underdense 
  environments, in $\S7$ we draw the conclusions. 
  
  A Hubble constant of $H_0$ = 70 km s$^{-1}$ Mpc$^{-1}$  is adopted throughout.
  
\section{The sample}

We have extracted the sample of elliptical galaxies from HyperLeda \cite{pa2003} requiring them to have 
$\sigma_0$ measured,  $m_{\rm B}$ $\le$ 15.5,  $v_{\rm r}$  $\in$ [2000 -- 10000]  km s$^{-1}$ ,
and $\delta \ge$ $0\,^{\circ}$ .  The constraints have been imposed to homogenize a sample that otherwise would not satisfy any predefined selection criteria,
having been  drawn from a compilation of available data.  The limit in $m_{\rm B}$ makes the sample to become flux limited,  the lower limit in $v_{\rm r}$ reduces distance uncertainty  due to the contribution of peculiar motions, the upper limit in $v_{\rm r}$  keeps 
contained the increase of luminosity with increasing distance and the limit in $\delta$ ensures homogeneous sampling of the environment for all ellipticals in the sample, 
since the catalog that we have used to estimate the environmnent (see next session) is limited to  $\delta \ge -2.5$ \,$^{\circ}$.

The selection based on the above described prescriptions produced a  
sample of 384 elliptical galaxies, that are listed in Table 1 \footnote{online only table.}.
For each elliptical, Table 1 reports identificator (column 1), equatorial coordinates $RA_{\rm J2000}$ and $Dec_{\rm J2000}$
(columns 2 and 3),
central velocity dispersion $\sigma_0$ and related uncertainty $\Delta\sigma_0$ (columns 4 and 5), total apparent B magnitude $m_{\rm B}$ and related uncertainty $\Delta$$m_{\rm B}$ (columns 6 and 7), radial
velocity  $v_{\rm r}$ corrected for the Virgo infall flow (column 8), number of neighbours $N_{\rm neigh}$ detected 
within the typical group/cluster scale (the details on the neighbour search method will be given in the next section).

Figure 1  shows the  distribution of our sample ellipticals (filled circles) in the $v_{\rm r}$, $M_{\rm B}$ plane together with the curve corresponding to the faintest observable  $M_{\rm B}$  in a sample limited to $m_{\rm B}$ = 15.5 (which is the limit that we have imposed to our sample). The curve allows one to visualize the well known effect induced by distance on 
 luminosity in a flux limited sample: since the minimum observable luminosity increases with distance, the farthest galaxies will be, on average,
  also the brightest ones. From Fig. 1 we see, however,  that the distance effect constrains  only the minimum observable luminosity  and that its real entity depends  quite strongly  on the galaxy distribution  
  in the   $v_{\rm r}$, $M_{\rm B}$ plane, which  will never be completely uniform, due the  intrinsic clumpiness of galaxy
  distribution (i.e. the presence of clusters and groups) and in the present case also to the possible  lack of some data which we cannot exclude as we are 
  dealing with a sample drawn from a compilation of available data. 
  
  The effect of distance on the luminosity can be perceived better from Fig. 2 which shows the absolute magnitude ($M_{\rm B}$) distribution 
  of ellipticals in the whole sample (dotted histogram) overimposed
  to the $M_{\rm B}$ distribution of ellipticals in 4 bins of increasing radial velocity, each spanning a 2000  km s$^{-1}$  range.
 Definition and characteristics of the bins are given in Table 2 providing for each bin (column 1),  radial velocity range (column 2), total number of ellipticals (column 3), mean (column 4)  
  and median (column 5)  value of the  $M_{\rm B}$ distribution (these latter to be compared with  -20.80 and -20.87, which are the corresponding values  of the whole sample).
  Figure 2 evidences rather clearly the progressive shift of the  $M_{\rm B}$ distribution of ellipticals as a function of the increasing distance and
  from columns 4 and 5 of Table 2 we see that this distance effect induces an average increase of 1.2 magnitudes between the farthest and the nearest ellipticals in our sample.


\section{The FJ relation for the whole sample}

The FJ relation for the total sample is shown in Fig.3. Superimposed to the data (open circles) is the best fit (green continuos line) derived taking into account errors on  
$\sigma_0$  (displayed in Fig. 3 as well) and corresponding to
$L_{\rm B}$ $\propto$ $\sigma_0^{4.0}$. The red dashed ($L_{\rm B}$ $\propto$ $\sigma_0^{5.6}$) and blue dotted ($L_{\rm B}$ $\propto$ $\sigma_0^{3.2}$) lines instead represent the best (weighted for errors on $\sigma_0$) fits for the 206 bright (Log($L_{\rm B}/L_{\rm B\odot})$   $\ge$ 10.5) 
and the 178 faint (Log($L_{\rm B}/L_{\rm B \odot})$ $<$ 10.5) ellipticals in the sample, with  Log($L_{\rm B}/L_{\rm B \odot})$ $=$ 10.5 being the value below which data in Fig. 3
start do deviate progressively from the fit and to extend towards low values of $\sigma_0$. 

From Fig. 3 we see that the dispersion of the data around the fits is quite large with the average scatter (RMS), $\sigma$($\sigma_0$), amounting 
to 51.1 $\pm$ 1.2 km s$^{-1}$ for the whole
 sample and to  49.3 $\pm$ 1.8 km s$^{-1}$ and 52.3 $\pm$ 1.7 km s$^{-1}$ respectively for the bright and faint subsamples.
  
 The less steep increase of  $L_{\rm B}$ with $\sigma_0$ displayed by  faint ellipticals in our sample confirms previous evidence obtained by several authors
 \cite{to1981,da1983,he1992,fr2005,ma2005,be2006,ds2007,la2007,vo2007,ko2012}
 on different samples which find theoretical justification \cite {de1986} in the expected link beween the FJ relation slope and the amount of dark matter in elliptical galaxies. 
 
 Some caution should be given however to slope values derived by means of fits in which errors on $\sigma_0$ (column 5 of Table 1) have been taken into account, as they 
 might result somewhat artificially increased if data with large $\sigma_0$  errors (which weight less) are mostly found below the fit lines. This seems actually 
 to be the case in our sample (cfr. Fig. 3) and from Fig. 4 we can visualize better the effect induced 
 by the weighted fit on the slopes. The distributions  of the relative  $\sigma_0$  errors ($\Delta$$\sigma_0$/$\sigma_0$)
 of data laying below (continuos) and above (dotted) the best fit lines for the unweighted (left) and weighted (right) fit of the whole sample (upper
 panels) and of the bright and faint subsamples (middle and lower panels) presented  in Fig. 4 show that
 the effect produced by the  weighted fit is to reduce the numerical dominance of ellipticals with small $\Delta$$\sigma_0$/$\sigma_0$ above the fit lines increasing, as a consequence, the slope of the FJ relation. Had we derived the slopes with an unweighted fit we would have obtained 
 $L_{\rm B}$ $\propto$ $\sigma_0^{3.4}$ for the whole sample,  
 $L_{\rm B}$ $\propto$ $\sigma_0^{4.7}$ for the bright  and  $L_{\rm B}$ $\propto$ $\sigma_0^{2.7}$ for the faint subsample (with average
 scatters $\sigma$($\sigma_0$) respectively equal to 51.9 $\pm$ 1.2 km s$^{-1}$, 49.5 $\pm$ 1.8 km s$^{-1}$ and 52.9 $\pm$ 1.7 km s$^{-1}$).
  
  It is worthwhile to stress that a weighted fit will always produce an artificial steepening of the slope (whatever the size of the errors) if data
   with large errors 
  are preferentially found below the fit. To prove that we have
 derived the FJ relation for the 314 ellipticals having  $\Delta$$\sigma_0$/$\sigma_0$  $\leq 0.1$, finding  that the weighted fit induces 
 an increases of the slope  from 3.8 to 4.1 for the whole sample, from  5.4 to 5.8 and from 3.0 to 3.3, respectively for the bright and faint subsample. The average dispersion of the data diminishes a little and
 sets around 48 km s$^{-1}$, the exact value depending on the kind of fit and sample.
 The slight reduction of $\sigma$($\sigma_0$)  is expected as we have excluded data with large relative errors on $\sigma_0$ (i.e. $\Delta$$\sigma_0$/$\sigma_0$  $>$ 0.1)
 which  being less accurate are more likely to deviate more strongly from the fit.
 
 Table 3 summarizes all the results described above. In column 1 we list the sample kind (whole, bright or faint), in column 2 the number
 of ellipticals in each sample, in column 3 the kind of fit (either unweighted or weighted), in column 4 the FJ relation slope ($\alpha$)
 with related uncertainty, in column 5 the average scatter of data around the best fit line ($\sigma$($\sigma_0$)) with related uncertainty.
 Figures in Table 3 do not allow us to establish the exact value for the slope of the FJ either of the whole sample or of its bright and faint
 components, but allow us to confirm the presence of two distinct (luminosity dependent) components in the FJ relation characterized 
 by a slope which is steeper for bright (Log($L_{\rm B}/L_{\rm B \odot})$ $\ge 10.5$) than for faint (Log($L_{\rm B}/L_{\rm B \odot})$ $<$ 10.5) elliptical, as 
 the difference between the slopes holds (and is larger than the errors)  whatever the kind of sample 
 (either whole or whole with small $\Delta$$\sigma_0$/$\sigma_0$) and of fit (either unweighted or weighted).

Finally, to check the effect of possible errors on $m_{\rm B}$ on the derived FJ relation, we have randomly added or subtracted to
 each  $m_{\rm B}$ (column 6 in Table 1) either its real (column 7 in Table 1) or  average error (computed on the whole sample). 
We have repeated this operation 300 times for both cases, thus obtaining 2 sets  of data each including 300 simulated samples.  We have subsequentely derived the FJ relation (weighted fit) for each simulated 
sample in each set and grouped  the results together, to obtain the total distribution of  $\sigma$($\sigma_0$),  $\alpha$, and $\alpha$$_{max}$- $\alpha$$_{min}$ (i.e. twice the maximum uncertainty  on the slope $\alpha$).
These distributions (normalized to the total number of simulated samples) are shown in Fig. 5, where 
 continuos and dotted histograms refer respectively  to samples  obtained by random addition or subtraction of the average or of the real error on $m_{\rm B}$.
Upper panels refer to the whole simulated samples, middle and lower panels to the bright and faint subsamples, extracted from the previous ones. Since the artificial
random increase or decrease of  $m_{\rm B}$ implies a decrease or increase in $L_{\rm B}$, the number of ellipticals in the high or low luminosity subsamples 
is not anymore constant but varies as a consequence of the variation in  $L_{\rm B}$. We find a median value of 204 and 180 ellipticals  and of 210 and 174 ellipticals (to be compared with the real value of 206 and 178) 
for the high and low luminosity subsamples  when the average or real error on $m_{\rm B}$ is respectively randomly added or subtracted. 

The arrow placed on each distribution of Fig. 5 indicates the corresponding value obtained in the real case (i.e. the weighted fit for 
the whole sample or for the bright or for the faint subsample, cfr. figures in Table 3, lines 4,5 and 6) and allow us to 
state that the 
effect of the error on  $m_{\rm B}$ is an increase in the average dispersion ($\sigma$($\sigma_0$)), a steepening of the slope  ($\alpha$)
and an increase of  the uncertainity of this latter quantity ($\alpha$$_{max}$- $\alpha$$_{min}$). From Fig. 5 we see that the effect of
the error on $m_{\rm B}$ is stronger for the whole sample than for the faint and bright subsamples, which
is not surprising as we have shown (in sect. 3) that the  FJ relation of the total sample can be interpreted as due to the combination of two distinguished
 (luminosity dependent) components. Variations in $L_{\rm B}$ of each
elliptical in the sample are thus expected to produce a stronger (amplified) effect on the FJ relation of the whole sample 
 than on the FJ relations of the separate (luminosity dependent) components. From Fig. 5 we also see  that the effect of the error
on $m_{\rm B}$ is stronger for the faint (lower panels) than for the bright (middle panels) subsample. This is 
not surprising too, since variations in $L_{\rm B}$ are expected to influence more strongly fits which have a less steep slope.

In Table 4 we list for each sample (column 1), the error $\Delta$$m_{\rm B}$, (either the average value for the whole sample or the range within which the true value is found)  which has been randomly added or 
subtracted to $m_{\rm B}$  (column 2), the mean value of the average scatter of the data around the fit $<$$\sigma$($\sigma_0$)$>$ and its RMS (columns 3 and 4), the mean value 
of the slope $<$$\alpha$$>$ and its
RMS (columns 5 and 6), the mean value of $\alpha$$_{max}$- $\alpha$$_{min}$ (here indicated as $<$$\Delta$$\alpha$$>$) and its RMS (columns 7 and 8).  Comparing the mean values listed in 
Table 4 (columns 3,5 and 7) with values obtained
for the real sample and subsamples (arrows in Fig. 5 corresponding to figures listed in Table 3, lines 4,5 and 6) we see that on average the
effect of the error on $m_{\rm B}$ can be considered moderate. Data in Table 3 and Fig. 5 allow us to conclude that even if the $m_{\rm B}$
of each elliptical would vary of a quantity respectively equal to the average or real error, which is surely an overestimate of what may
happen in the real case, we could however confirm the value of the average dispersion of the data ($\sigma$($\sigma_0$)) around the best fit
which would get only slightly increased. The presence of two distinct (luminosity dependent) components would be confirmed as well, since both
slopes would get somewhat increased but remain still well distinguished (their difference being larger than their errors).


\section{The environmnent}

To evaluate the environment of each elliptical galaxy  in our sample we have applied the neighbour search code 
of  Focardi \& Kelm (2002)  
to the Updated Zwicky Catalog (UZC) \cite{fa1999}. 

UZC is a wide angle 3D catalog of  nearby galaxies that 
covers the entire northern sky down to a declination of -2.5\,$^{\circ}$, 
and  is claimed  \cite{fa1999}  to be  96\% complete for galaxies brighter than $m_{\rm B}$ = 15.5. 

The neighbour search code is a versatile tool which can be applied to 3D catalogs 
either to produce galaxy samples characterized by different environment  or to estimate 
galaxy environment on different scales and  depth.

 Detailed description of the code can be found in  Focardi \& Kelm (2002), together with the results of the  first application
of the code  to UZC which has produced
 a large homogeneous sample of compact groups (UZC-CGs, Focardi \& Kelm 2002). The code has been subsequentely applied to UZC to produce a sample of bright isolated 
galaxy pairs (UZC-BPGs, Focardi et al. 2006) and a small sample of 
very isolated bright ellipticals \cite{me2009}. It has also been applied to the 2dFGRS \cite{co2001} in order to perform a detailed analsys on
the luminosity/environment/spectral type relation for galaxies \cite{ke2005}.  

When used simply to detect neighbours, as in the present case,  the code needs only two input parameters which are the maximum projected distance ($\Delta R$)  
and radial velocity difference  ($|$$\Delta v_{\rm r}$$|$) 
between each elliptical in the sample  and its  possible neighbours (from UZC)  and includes  obviously, a cross check on coordinates, $v_{\rm r}$  and $m_{\rm B}$ to avoid spurious 
detection of the elliptical 
as possible neighbour of itself.   
 
We have set $\Delta R$ = 1.5  Mpc and $|$$\Delta v_{\rm r}$$|$ = 1000  km s$^{-1}$, a choice which has allowed us to estimate
the environment on the typical scale of galaxy clusters.

 Figure 6 shows the distribution of the number of neighbours ($N_{\rm neigh}$) when the whole sample is divided into the 4 bins
 of increasing radial velocity, that we have defined  in $\S$2 and whose characteristics are summarized in Table 2.
 Distributions appear rather different, which is not unexpected as we have already shown (cfr. Figs. 1 and  2,  Table 2 columns 4 and 5)
 how  distance effect produces  an  increase of  luminosity with increasing distance. Thus, as 
 the number of galaxies decreases with increasing luminosity we would expect, on average, a decreasing number of neighbours with increasing distance. Figure 6 shows that
 this is not exactly the case as the  expected decrease in $N_{\rm neigh}$ is evident only in the fourth bin, while both bins II and III show an  anomalous tail (large values of $N_{\rm neigh}$),
 which is due to the presence of several galaxy clusters and groups (belonging respectively  the Perseus-Pisces and Coma supercluster) 
 of which some  ellipticals in bins II and III are members.


\section{The effect of the environment on the FJ relation}

 To look for possible effects induced by the environment on the FJ relation one must compare  
 ellipticals in  
 high and 
 low density environments, with density being as large and as small as possible
 but leaving however a number of ellipticals
 in each environment that is not too small.
 
An objective way to define these extreme environments can be obtained relating the number of neighbours requested to enter each sample to the median value
of the  $N_{\rm neigh}$  distribution, this latter computed separately for each bin to account for effects related to distance and non uniformity in the galaxy distribution. 

We find that defining as high or low density an environment characterized by a number of neighbours equal or larger or equal or smaller than 5 times or 0.2 times the median value of
$N_{\rm neigh}$ provides us with two subsamples of 26 and 36 ellipticals (respectively including 7\% and 9\% of the whole sample). Since the median value of  $N_{\rm neigh}$ is equal to 8 in first bin,
to 9 in the second and third bin and to 4 in the fourth bin,  our definition of high density environmnent implies $N_{\rm neigh}$ $\ge 40$ (in bin I),   $N_{\rm neigh}$ $\ge 45$ (in bins II and III)
 and  $N_{\rm neigh}$ $\ge 20$ (in bin IV),
while our definition of low density environment implies  $N_{\rm neigh}$ $\le 1$ (in bins I,II and III) and   $N_{\rm neigh}$ = 0 (in bin IV).


 Figure 7 shows the FJ relation for ellipticals in the high (red filled circles) and  low (blue open squares) density environments defined above.
 Each data is displayed with its own error (both in $m_{\rm B}$ and in $\sigma_0$), while  the red continuos and blue dotted lines represent the best weighted fits respectively giving
 $L_{\rm B}$ $\propto$ $\sigma_0^{3.79}$ and $L_{\rm B}$ $\propto$ $\sigma_0^{3.75}$.
 It is evident that ellipticals in  low density  regions  distribute in a much more dispersed way than ellipticals in high density
 ones. The $\sigma$($\sigma_0$) of the FJ in low density environments  (67.0 km s$^{-1}$) is in fact almost twice the one found
 in high density  ones  (33.7 km s$^{-1}$) and even larger than what found for the whole sample (51.1 km s$^{-1}$).

 Weighted and unweighted fit give exactly the same slope  ($\alpha$ $\simeq$ 3.8) in high density environments, while in low density environments the slope is steeper
 in the weighted fit ($\alpha$  $\simeq$ 3.8) than in the unweighted one ($\alpha$ $\simeq$ 3.3), due to the effect produced by the 
 dominance of data with larger  $\sigma_0$ errors below the
 fit line (as discussed in sect. 3).
 
 One might then argue that the large $\sigma$($\sigma_0$) displayed by ellipticals in low density environments could be induced from their large  $\sigma_0$ errors, but if we exclude from the sample the 6 ellipticals with 
 the largest relative errors ($\Delta$$\sigma_0$/$\sigma_0$  $\geq 0.15$)
 we still get large values for $\sigma$($\sigma_0$) (67.9   km s$^{-1}$ and  68.5  km s$^{-1}$  respectively for the weighted and unweighted fit).
 Moreover, from Fig. 7 we see that the luminosity distribution of ellipticals in high and low density environments is rather
 similar, implying that the larger value of  $\sigma$($\sigma_0$) displayed by ellipticals in the latter sample cannot be attributed to a
 luminosity effect linking  intrinsic scatter to luminosity, as already evidenced for the FP (see e.g. Bender et al. 1992; Hyde \& Bernardi 2009). 
 The fraction of faint (Log($L_{\rm B}/L_{\rm B \odot})$ $<$ 10.5) ellipticals is in fact almost the same in high (15/26,  
 $\simeq$ 58 \%) and low (20/36, $\simeq$ 56 \%) density environments, implying that the larger $\sigma$($\sigma_0$) displayed by ellipticals in low density environmnent 
 cannot be attributed to a larger content of low luminosity ellipticals.
 
 An accurate comparison of the luminosity distribution of ellipticals in high and low density environments can be obtained 
 from Fig. 8, which shows the normalized $M_{\rm B}$ distribution of ellipticals of the two samples.
 From Fig. 8 we see that ellipticals in high density
  environments (shaded histogram) cover a 
 somewhat larger luminosity range, extending 
 at both  sides of the distribution while  ellipticals in low density environments 
 dominate at  $ M_{\rm B}$ $\sim -21$. The median value of the distributions is exactly the same ($M_{\rm B}$ = -20.63) and 
 is indicated by an arrow  in Fig. 8 and the difference between the distributions is not at all significant, as confirmed by the KS test which gives a probability of p=0.77
 that the two distributions are similar.
 However, if we eliminate 5 ellipticals (the 4 brightest  and the faintest one) in the high density sample and the 2 faintest ellipticals in the 
 low density one, so as to make both samples  to cover exactly  the same range in luminosity, 
we find $\alpha$ = 3.7,  $\sigma$($\sigma_0$)= 32.1 km s$^{-1}$ for the unweighted fit and $\alpha$ = 3.9,  $\sigma$($\sigma_0$)= 32.2 km s$^{-1}$, for the weighted fit in the high density sample,
$\alpha$ = 2.7,  $\sigma$($\sigma_0$)= 69.9 km s$^{-1}$ for the unweighted fit and $\alpha$ = 3.6, $\sigma$($\sigma_0$)= 68.6 km s$^{-1}$ 
for the weighted fit in the low density sample, confirming what we have obtained on the whole luminosity range.

Finally to check how solid can be considered our result we have inspected the SDSS-III (DR 8) database \cite{yo2000,ai2011}. looking for 
$\sigma_0$ measures for ellipticals in the low and high density environments. Unfortunately those data are
available only for 14 ellipticals in the high density environment and for 12 ellipticals in  the low density one and are reported in
Tables  5 and 6, in which we list for each elliptical in each sample, identificator (column 1), $\sigma_0$ value from the SDDS (when available) with related uncertainty (column 2) and
the difference ($\Delta\sigma_0$) between  SDSS  and  Hyperleda value for  $\sigma_0$ (column 3).  Inspection of data in column 3  reveals a  $\Delta\sigma_0$
which is in general small, but  almost always  negative in the high density environment ($<$$\sigma_0$$>$ = -13.2 km s$^{-1}$,
$<$$\sigma_0$$>$$_{\rm RMS}$ = 9.8 km s$^{-1}$)  and that is larger and more spread around the zero in low density environment
($<$$\sigma_0$$>$ = 5.9 km s$^{-1}$, $<$$\sigma_0$$>$$_{\rm RMS}$ = 24.2 km s$^{-1}$). 

Replacing SDSS $\sigma_0$ (when available) to Hyperleda ones gives $L_{\rm B}$ $\propto$ $\sigma_0^{3.8}$ for the high density environment (in both weighted and
unweighted fit) and $L_{\rm B}$ $\propto$ $\sigma_0^{4.3}$, $L_{\rm B}$ $\propto$ $\sigma_0^{3.4}$ for the low density environment (respectively for the weighted and and unweighted fit).
Average scatter of the data  gets somewhat worst, especially in high density environments. We find   $\sigma$($\sigma_0$) = 36.5 km s$^{-1}$ for the high density enironments and 
 $\sigma$($\sigma_0$) = 68.1 km s$^{-1}$ for the low density ones (to be compared with 33.7 km s$^{-1}$ and 67.0 km s$^{-1}$ which are the corresponding values obtained  for the weighted fits when using only
 data from Hyperleda). 
 Despite of the slight increase of  $\sigma$($\sigma_0$) which is likely to be due to the non homogeneity between the two
 distinct sets of data (as proved by the shift in $\sigma_0$ reported in Tables 5 and 6, column 3), the environment effect on $\sigma$($\sigma_0$) is confirmed. 
 
 Table 7 summarizes all the results described above, in column 1 we indicate the sample kind (overdense and underdense stand for
 high density and low density) and
 within parenthesis its caractheristics  related to $\sigma_0$ (either small error or data from the SDSS), in column 2
 the sample size, in column 3 the kind of fit (either unweighted or weighted), in column 4 the FJ relation slope
 and related uncertainty, in column 5 the average dispersion of the data around the fit.
  
  Finally, in analogy with what we have done for the whole sample (see sect. 3, Fig. 5 and Table 4)  we have checked for possible maximum effects due to $m_{\rm B}$
  errors on the FJ relation in the high and low density environment. The procedure is exactly the same but in this case we have generated only 30 simulated samples
  by random addiction or subtraction of the real error on  $m_{\rm B}$. The results are shown in Fig. 9 showing the normalized distribution
  of $\sigma$($\sigma_0$), $\alpha$  and $\Delta\alpha$ for the high density (upper panels)  and low density  (lower panels) simulated samples. 
  The arrow on each plot indicates the value obtained in the real case.
  From Fig 9
  we see that errors on luminosity would produce a general degradation of the fit quality (particularly evident in the 
  possible large increase of $\alpha$ and  $\Delta\alpha$ for the low density sample), but that however the difference in  
  $\sigma$($\sigma_0$) would be mantained.

\section{ How much overdense and underdense have to be the environments?}

In the previous section we have shown that ellipticals in  high density environments display a significant reduction of the FJ  scatter, when
compared to ellipticals in low density ones. Both kind of environments have been selected objectively requiring a number of neighbours  ($N_{neigh}$)
equal or larger or equal or smaller than 5 times or 0.2 times the median value of the $N_{neigh}$ distribution (computed separately for each distance bin).

We now reduce and increase progressively the overdensity and underdensity value (i.e. the multiplicative factor that we have applied to the median value of  $N_{\rm neigh}$) 
to check the level of densities at which the difference in the FJ scatter holds.
 
 The results of this test are shown in Tables 8  and 9 where for each value of the overdensity or underdensity (column 1) we indicate the number of ellipticals in each sample (column 2),
  the kind of fit (column 3), the slope with its uncertainty (column 4) and  
 the  average scatter  $\sigma$($\sigma_0$) of the FJ relation with related uncertainty(column 5). Tables 8 and 9 allow one to follow the increase of 
 $\sigma$($\sigma_0$), as overdense environment becomes less and less dense and, 
 complementary, the decrease of  
 $\sigma$($\sigma_0$), as underdense environment gets more and more dense. From Table 8 we see that $\sigma$ ($\sigma_0$) mantains its small value for  overdensities down to a value of 3.5, that it is still 
 small, even if somewhat increased, when  overdensity is equal to 3 and that then it starts to increase more rapidly to reach the characteristic value displayed by the whole sample at an overdensity of 1.5. 
 From Table 9 instead we see that the dispersion is already below 60  km s$^{-1}$ at an underdensity factor of 0.25 and that it decreases progressively remaining just above the 
 $\sigma$($\sigma_0$) of the whole sample when the underdensity factor is equal to 0.75.
 
 This progressive increase/decrease of $\sigma$ ($\sigma_0$) with decreasing/increasing density in overdense/underdense environments 
 gives more strength to our result confirming an effect relating environment to the FJ relation scatter.

 \section{Conclusions}
 
 Using a sample of 384 nearby elliptical galaxies
  and objectively estimating their environment on the basis of the number of neighbours
 within the typical galaxy cluster and group scale we have provided evidence for
 an effect relating 
 the intrinsic scatter of the FJ relation to the environment. We have shown that the scatter of the FJ is reduced to almost half of its value when ellipticals
 in highest overdensities are compared to ellipticals in less-density environments, that the effect
 is not induced by luminosity differences between the samples and that it holds for overdensities 
 ranging between 3.5 and 5 the median value
 of the number of neighbours distribution.
 Besides indicating a rather simple and quite natural way to reduce the large scatter affecting the FJ relation, our result, if confirmed on
 larger samples, is 
 very likely to open  an interesting perspective for models
 of galaxy formation.
 
\acknowledgments

We would like to thank the anonymous referee, for his/her constructive comments and suggestions which helped us to greatly improve the scientific content of this paper.

We acknowledge the usage of HyperLeda database (http://leda.univ-lyon1.fr) and of SDSS-III.

Funding for SDSS-III has been provided by the Alfred P. Sloan Foundation, the Participating Institutions, the National Science Foundation, and 
the U.S. Department of Energy Office of Science. The SDSS-III web site is http://www.sdss3.org/.

SDSS-III is managed by the Astrophysical Research Consortium for the Participating Institutions of the SDSS-III Collaboration including the 
University of Arizona, the Brazilian Participation Group, Brookhaven National Laboratory, University of Cambridge, Carnegie Mellon University, University of Florida, 
the French Participation Group, the German Participation Group, Harvard University, the Instituto de Astrofisica de Canarias, the Michigan State/Notre Dame/JINA Participation Group, 
Johns Hopkins University, Lawrence Berkeley National Laboratory, Max Planck Institute for Astrophysics, Max Planck Institute for Extraterrestrial Physics, New Mexico State University,
 New York University, Ohio State University, Pennsylvania State University, University of Portsmouth, Princeton University, the Spanish Participation Group, University of Tokyo, 
 University of Utah, Vanderbilt University, University of Virginia, University of Washington, and Yale University.

This work has been  supported by MIUR.

\clearpage

%
%
%

\begin{table*}
\begin{center}
\begin{tabular}{ccrcc}
\tableline
\tableline
Bin & $\Delta$ $v_{\rm r}$ & $N_{\rm ell.}$ & $<$$M_{\rm B}$$>$ & med($M_{\rm B})$     \\
 & km s$^{-1}$ & & \\
\tableline
 I      & 2000 $\le$ $v_{\rm r}$ $<$ 4000   & 61 & -20.25 & -20.21 \\
 II     & 4000 $\le$ $v_{\rm r}$  $<$ 6000   & 136 & -20.59& -20.67    \\
 III    & 6000 $\le$ $v_{\rm r}$  $<$ 8000   & 120 &-20.98  &-21.02  \\
 IV  &  8000 $\le$ $v_{\rm r}$  $\le$ 10000   &	67 & -21.39  &-21.41 \\
 \tableline
\end{tabular}
\end{center}
\tablenum{2}
\caption{The effect of distance on luminosity on our sample. \label{tbl-2}}
\end{table*}

\clearpage

\begin{table*}
\begin{center}
\begin{tabular}{lllcl}
\tableline
\tableline
Sample & $N_{\rm ell.}$ & fit &  $\alpha$ & $\sigma$($\sigma_0$)\\
& & &  & km s$^{-1}$ \\
\tableline
 whole      & 384  & unweighted & 3.4$^{+0.3}_ {-0.1}$ &  51.9 $\pm 1.2$  \\
 bright    & 206  & unweighted & 4.7$^{+0.8}_ {-0.6}$ & 49.5 $\pm$ 1.8  \\
faint    & 178 & unweighted & 2.7$^{+0.5}_{-0.3}$ &  52.9 $\pm$ 1.7 \\
  whole      & 384  & weighted & 4.0$^{+0.2}_{-0.2}$ &  51.1 $\pm$ 1.2  \\
  bright    & 206  & weighted & 5.6$^{+0.9}_{-0.7}$ & 49.3 $\pm$ 1.8  \\
  faint   & 178  & weighted & 3.2$^{+0.5}_{-0.4}$ &  52.3 $\pm$ 1.7   \\
   whole ($\Delta\sigma_0/\sigma_0$ $\leq 0.1$) & 314 & unweighted & 3.8$^{+0.2}_{-0.2}$ &   48.4 $\pm$ 1.0 \\
    bright ($\Delta\sigma_0/\sigma_0$ $\leq 0.1$) & 174 & unweighted &  5.4$^{+1.0}_{-0.8}$ & 47.9 $\pm$ 1.5 \\
     faint ($\Delta\sigma_0/\sigma_0$  $\leq 0.1$)  & 140  & unweighted &  3.0$^{+0.5}_{-0.3}$ &  47.2 $\pm$ 1.1 \\
  whole ($\Delta\sigma_0/\sigma_0$ $\leq 0.1$) & 314 & weighted & 4.1$^{+0.3}_{-0.2}$ &   48.1 $\pm$ 1.0 \\
   bright  ($\Delta\sigma_0/\sigma_0$  $\leq 0.1$)   & 174 & weighted &  5.8$^{+1.1}_{-0.8}$ & 48.0 $\pm$ 1.5 \\
 faint ($\Delta\sigma_0/\sigma_0$  $\leq 0.1$) & 140 & weighted &  3.3$^{+0.6}_{-0.4}$ &  47.0 $\pm$ 1.1\\

 \tableline
\end{tabular}
\end{center}
\tablenum{3}
\caption{FJ relation parameters for ellipticals in the whole sample and in the bright (Log($L_{\rm B}/L_{\rm B\odot})$   $\ge$ 10.5)  
and faint  ( Log($L_{\rm B}/L_{\rm B \odot})$ $<$ 10.5  ) subsamples. \label{tbl-3}}
\end{table*}

\clearpage

\begin{table*}
\begin{center}
\begin{tabular}{lcccccll}
\tableline
\tableline

Sample & $\Delta m_B$  & $<$$\sigma$($\sigma_0$)$>$ & $\sigma$($\sigma_0$)$_{RMS}$ &  $<$ $\alpha$ $>$ & $\alpha$$_{RMS}$  & $<$ $\Delta\alpha$$>$  & $\Delta\alpha$$_{RMS}$\\
&   &  km s$^{-1}$ & km s$^{-1}$ &  & & &  \\
\tableline
 whole    &  0.25  &  52.6  & 0.6 &  4.3 & 0.1 & 0.52 & 0.03 \\
 whole     & [0.03 -- 0.72]    &  52.8  & 0.7 &  4.3 & 0.1  & 0.52 & 0.03 \\
 bright  & 0.25   &   49.8 & 1.2 &  5.9 & 0.5 &  1.7 & 0.3 \\ 
 bright  &  [0.03 -- 0.72]   &   50.1 & 1.3 &  5.7 & 0.5 &  1.6 & 0.3 \\ 
 faint  &  0.25  &   54.5 & 1.5 &  3.5 & 0.3 &  1.1 & 0.2 \\ 
  faint   & [0.03 -- 0.72]  &   54.2 & 1.8 &  3.7 & 0.3 &  1.2 & 0.3 \\

 \tableline
\end{tabular}
\end{center}
\tablenum{4}
\caption{Average values for the FJ relation parameters derived accounting
for possible errors on $m_{\rm B}$. \label{tbl-4}}
\end{table*}

\clearpage

\begin{table*}
\begin{center}
\begin{tabular}{lcr}
\tableline
\tableline
Identificator  & $\sigma_0$(SDSS)  & $\Delta\sigma_0$ \\
 &    km s$^{-1}$ & km s$^{-1}$   \\
\tableline 
 PGC 6945 &  -    & -\\ 
 NGC 704   & -   & -  \\ 
 NGC 3837   & -  &  - \\ 
 NGC 3842 & 291 $\pm 5$ & -24 \\
 NGC 3862  & 260 $\pm 5$  & -11 \\  
 NGC 4261 & - & - \\
 NGC 4473 & - & - \\
 NGC 4816 & - & -\\
 PGC 44137  & -   &  - \\ 
  NGC 4839 & 269 $\pm 5$ & -16 \\
 NGC 4842A & 208 $\pm 4$ & -8 \\
 PGC 44367 & 144 $\pm  3$ & -12 \\ 
 PGC 44467   & -  &  - \\ 
 NGC 4860 &  263 $\pm 4$ & -13 \\
 IC 3959    & 205 $\pm 4$ &  -3 \\
 NGC 4864      & -  & -\\   
 NGC 4869 & - &  - \\
 NGC 4874 & - &  - \\
 NGC 4881   & 193 $\pm 3$ & -7 \\
 NGC 4882 & 149 $\pm$ 3 & -14 \\
 NGC 4884  & - &  - \\
 NGC 4906   & 173  $\pm 3$ & 1  \\

  \tableline
                                  
 \end{tabular}                              
\end{center}
\tablenum{5}
\caption{Ellipticals in the overdense environment \label{tbl-5}}
                                
\end{table*}

\clearpage

\begin{table*}
\begin{center}
\begin{tabular}{lcr}
\tableline
\tableline
Identificator  & $\sigma_0$(SDSS)  & $\Delta\sigma_0$ \\
 &    km s$^{-1}$ & km s$^{-1}$   \\
\tableline 
IC 4041 & 119 $\pm 3$ & -17 \\
 IC 4045 & 208 $\pm 3$ & -9 \\
 PGC 44848 & 173 $\pm 3$ & -40\\
 NGC 4926 & 264 $\pm 4$ & -12 \\

\tableline
                                  
 \end{tabular}                              
\end{center}
\tablenum{5}
\caption{continued. \label{tbl-5}}
                                
\end{table*}

\clearpage

\begin{table*}
\begin{center}
\begin{tabular}{lcr}
\tableline
\tableline
Identificator  & $\sigma_0$(SDSS)  & $\Delta\sigma_0$ \\
 &    km s$^{-1}$ & km s$^{-1}$   \\
\tableline 
  NGC 631  	& -  & -\\ 
   NGC 810   &  292  $\pm 5$  & 34\\ 
  NGC 1226     &  - & -  \\ 
  UGC 3549  & - & -  \\  
 UGC 3844   & 189  $\pm 3$  & 39\\ 
 NGC 2474   & - &  - \\ 
 NGC 2800     & -  & -  \\ 
 NGC 2954  &   188  $\pm  3$ & -28 \\
   UGC 5313  &	63  $\pm 4$  & -11\\   
  IC 590   & 296  $\pm 5$ & 23 \\
    NGC 3392  & 165 $\pm 2$ & 6  \\
  NGC 3731 & 154  $\pm 3$ & -19 \\
 NGC 4187  & 277 $\pm 5$ & -23 \\
 NGC 4272  & - & -   \\
 UGC 7813  & 256 $\pm 4$ & -16 \\
  NGC 5583  & - & -\\
  NGC 5628 & 254  $\pm 4$ & 29 \\
    NGC 5771  & 122 $\pm 2$ & 22\\
IC 1101    &  -  & -  \\
 NGC 6020 & 205 $\pm 3$ & 15 \\
 NGC 6051 &  - & -  \\
 IC 1211    &  - &  - \\

 \tableline
                                  
 \end{tabular}                              
\end{center}
\tablenum{6}
\caption{Ellipticals in the underdense environment \label{tbl-6}}
                                
\end{table*} 
\clearpage

\begin{table*}
\begin{center}
\begin{tabular}{lcr}
\tableline
\tableline
Identificator  & $\sigma_0$(SDSS)  & $\Delta\sigma_0$ \\
 &    km s$^{-1}$ & km s$^{-1}$   \\
\tableline          
  NGC 6442 & - & -  \\
 NGC 6515     & - &  - \\
  NGC 6575  & - & -\\
 NGC 6697    & - & -\\   
  NGC 6702   &  -  & - \\
 IC 1317     & - &  - \\
 NGC 7052   &  -  & -  \\
 NGC 7360  &  -  & -  \\
 NGC 7512    & - & -  \\
 PGC 71599  & - &  - \\
 NGC 7735   &  -  & - \\
 NGC 7751 &  -  & -   \\
 NGC 7785  &  -  & -   \\
 NGC 7786   &  -  & - \\

 \tableline
                                  
 \end{tabular}                              
\end{center}
\tablenum{6}
\caption{continued. \label{tbl-6}}
                                
\end{table*}             

\clearpage

\begin{table*}
\begin{center}
\begin{tabular}{lclcc}
\tableline
\tableline
Sample & $N_{\rm ell.}$ & fit & $\alpha$ & $\sigma$($\sigma_0$)\\
&  &  & & km s$^{-1}$ \\
\tableline   
overdense  & 26 & unweighted & 3.8$^{+0.5}_{-0.4}$ & 33.7 $\pm$ 1.4\\
overdense & 26 & weighted &  3.8$^{+0.5}_{-0.4}$ & 33.7 $\pm$ 1.4\\
overdense (with 14 SDSS $\sigma_0$) & 26    & unweighted  & 3.8$^{+0.6}_{-0.5}$ & 36.6 $\pm$ 0.9 \\
overdense (with 14 SDSS $\sigma_0$) & 26 & weigthted  & 3.8$^{+0.6}_{-0.5}$ & 36.5 $\pm$ 0.9 \\
 underdense  & 36 & unweighted&  3.3$^{+1.4}_{-0.7}$ & 67.7 $\pm$ 3.7\\
 underdense & 36 & weighted & 3.8$^{+1.2}_{-1.1}$ & 67.0  $\pm$ 3.7\\
 underdense ($\Delta\sigma_0/\sigma_0$ $\leq 0.15$) & 30 & unweighted & 2.4$^{+0.7}_{-0.5}$ & 68.5 $\pm$ 2.9 \\
 underdense ($\Delta\sigma_0/\sigma_0$ $\leq 0.15$) & 30 & weighted & 3.4$^{+1.3}_{-1.0}$ & 67.9 $\pm$ 2.9 \\
 underdense (with 12 SDSS $\sigma_0$) & 36 & unweighted  & 3.4$^{+1.5}_{-0.8}$ & 69.0 $\pm$ 2.7 \\
 underdense (with 12 SDSS $\sigma_0$) & 36   & weighted & 4.3$^{+2.1}_{-1.0}$ & 68.1 $\pm$ 2.6 \\
  
 \tableline
\end{tabular}
\end{center}
\tablenum{7}
\caption{The FJ relation in overdense and underdense environments. \label{tbl-7}}
\end{table*}

\clearpage

 \begin{table*}
\begin{center}
\begin{tabular}{crlcc}
\tableline
\tableline
overdensity & $N_{\rm ell}$ & fit & $\alpha$ & $\sigma$($\sigma_0$) \\
 &   & & & km s$^{-1}$ \\
\tableline
 4.5    &  37  &   unweighted & 3.6$^{+0.4}_{-0.3}$ & 32.0 $\pm$ 1.4 \\
 4.5   & 37 & weighted & 3.7$^{+0.4}_{-0.3}$ 	& 32.0   $\pm$ 1.3 \\
 4.0    & 45 & unweighted & 3.7$^{+0.4}_{-0.3}$ 	& 34.1  $\pm$ 1.2      \\
 4.0 & 45 & weighted & 3.8$^{+0.5}_{-0.3}$  &  34.0  $\pm$ 1.2 \\
 3.5 & 56 & unweighted & 3.6$^{+0.3}_{-0.3}$ 	& 33.2  $\pm$ 1.1 \\
 3.5 & 56 & weighted & 3.8$^{+0.4}_{-0.3}$ 	& 33.0  $\pm$ 1.1 \\
 3.0  & 73  & unweighted & 3.6$^{+0.3}_{-0.3}$ & 35.6 $\pm$ 1.2\\
 3.0 & 73 & weighted & 3.8$^{+0.4}_{-0.2}$ & 35.0 $\pm$ 1.2\\
 2.5 & 88 & unweighted & 3.8$^{+0.3}_{-0.3}$ &  41.1 $\pm$ 1.1\\
 2.5 & 88 & weighted & 4.0$^{+0.4}_{-0.3}$ & 40.7\ $\pm$ 1.1\\
 2.0 & 108 & unweighted & 3.9$^{+0.4}_{-0.3}$ &  45.6 $\pm$ 1.9  \\
 2.0 & 108 & weighted & 4.1$^{+0.3}_{-0.4}$ &  45.2 $\pm$ 1.9 \\
 1.5 & 138 & unweighted & 3.8$^{+0.4}_{-0.3}$ &  51.4 $\pm$ 1.8 \\
 1.5 & 138 & weighted & 4.1$^{+0.4}_{-0.3}$ &  50.9 $\pm$ 1.8 \\
 1.0 & 198 & unweighted & 3.6$^{+0.3}_{-0.2}$ &  51.1 $\pm$ 1.5  \\
 1.0 & 198 & weighted & 4.1$^{+0.4}_{-0.2}$ &  50.2 $\pm$ 1.5 \\
 
 \tableline
                                  
 \end{tabular}                              
\end{center}
\tablenum{8}
\caption{Relaxing the overdensity.\label{tbl-8}}
                                
\end{table*}

\clearpage

\begin{table*}
\begin{center}
\begin{tabular}{crlcc}
\tableline
\tableline
underdensity & $N_{\rm ell}$ & fit & $\alpha$ & $\sigma$($\sigma_0$) \\
 &   & & & km s$^{-1}$ \\
\tableline
 0.25    &  78  &   unweighted & 3.4$^{+0.7}_{-0.5}$ & 59.3 $\pm$ 2.6  \\
 0.25   & 78 & weighted & 3.9$^{+0.9}_{-0.7}$ 	& 58.8 $\pm$ 2.6    \\
 0.40    & 99 & unweighted & 3.5$^{+0.7}_{-0.4}$ 	& 58.1 $\pm$ 2.4      \\
 0.40 & 99 & weighted & 3.9$^{+0.8}_{-0.5}$  &  57.6 $\pm$ 2.4 \\
 0.50 & 127 & unweighted & 3.1$^{+0.4}_{-0.3}$ 	& 55.9 $\pm$ 2.6   \\
 0.50 & 127 & weighted & 3.6$^{+0.5}_{-0.4}$ 	& 55.2 $\pm$ 2.5  \\
 0.75  & 158  & unweighted & 2.9$^{+0.3}_{-0.2}$ & 53.8$\pm$ 2.2 \\
 0.75 & 158 & weighted & 3.4$^{+0.4}_{-0.3}$ & 53.2 $\pm$ 2.2\\

 \tableline
                                  
 \end{tabular}                              
\end{center}
\tablenum{9}
\caption{Relaxing the underdensity.\label{tbl-9}}
                                
\end{table*}

\clearpage

\figcaption[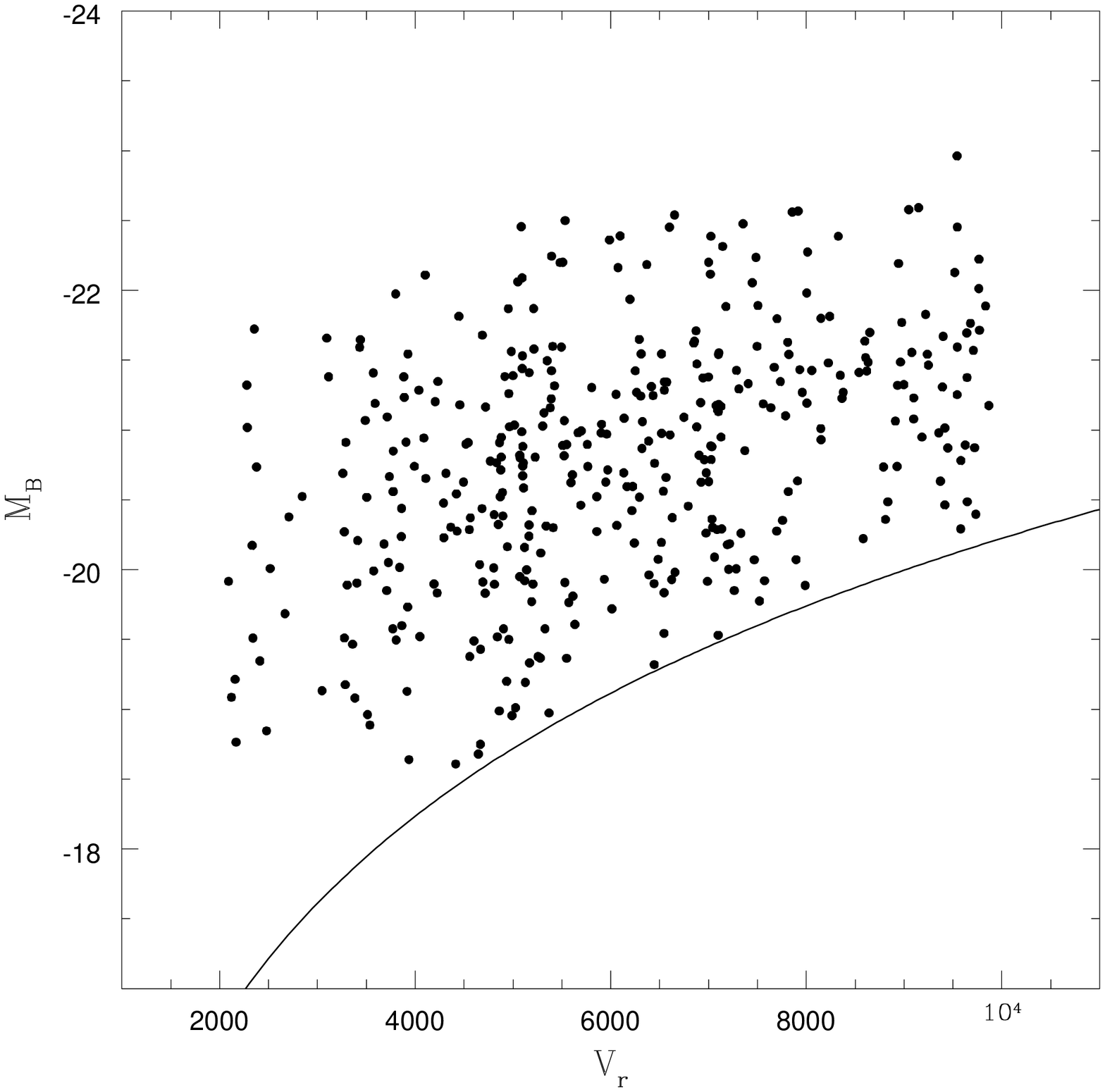]{The distribution of the whole sample ellipticals (filled circles)  in the  v$_r$, $M_{\rm B}$ plane. All data lay above the curve representing the increase of the minimum observable luminosity  ($M_{\rm B}$) with distance for
galaxies having m$_B$ $\le 15.5$. (This latter is the limit that we have imposed to the sample). \label{fig1}}

\figcaption[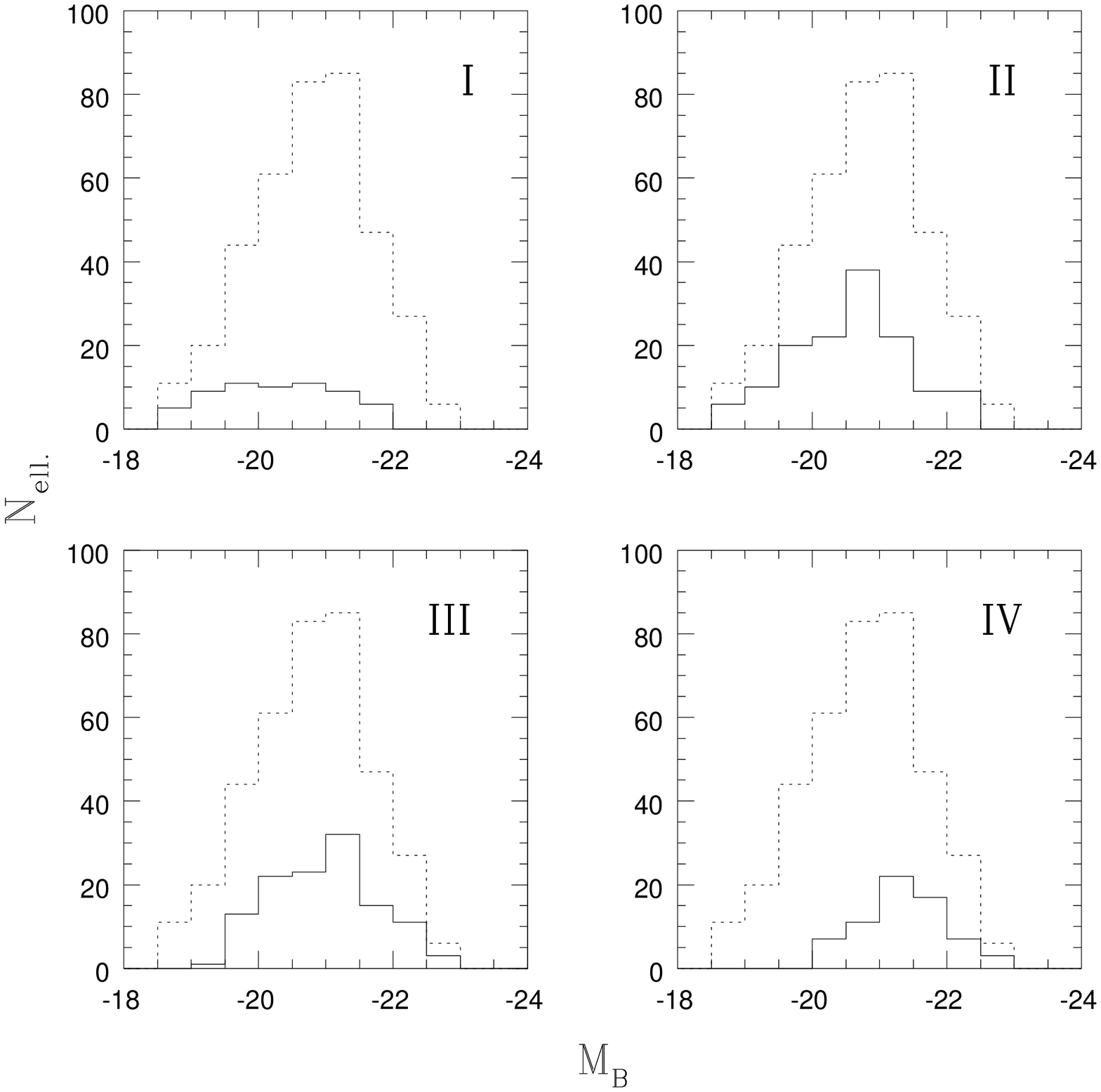]{Relative contribution to the absolute blue magnitude ($M_{\rm B}$) distribution of the whole sample (dotted) from ellipticals belonging to 4 bins of increasing distance (see Table 2). \label{fig2}}

\figcaption[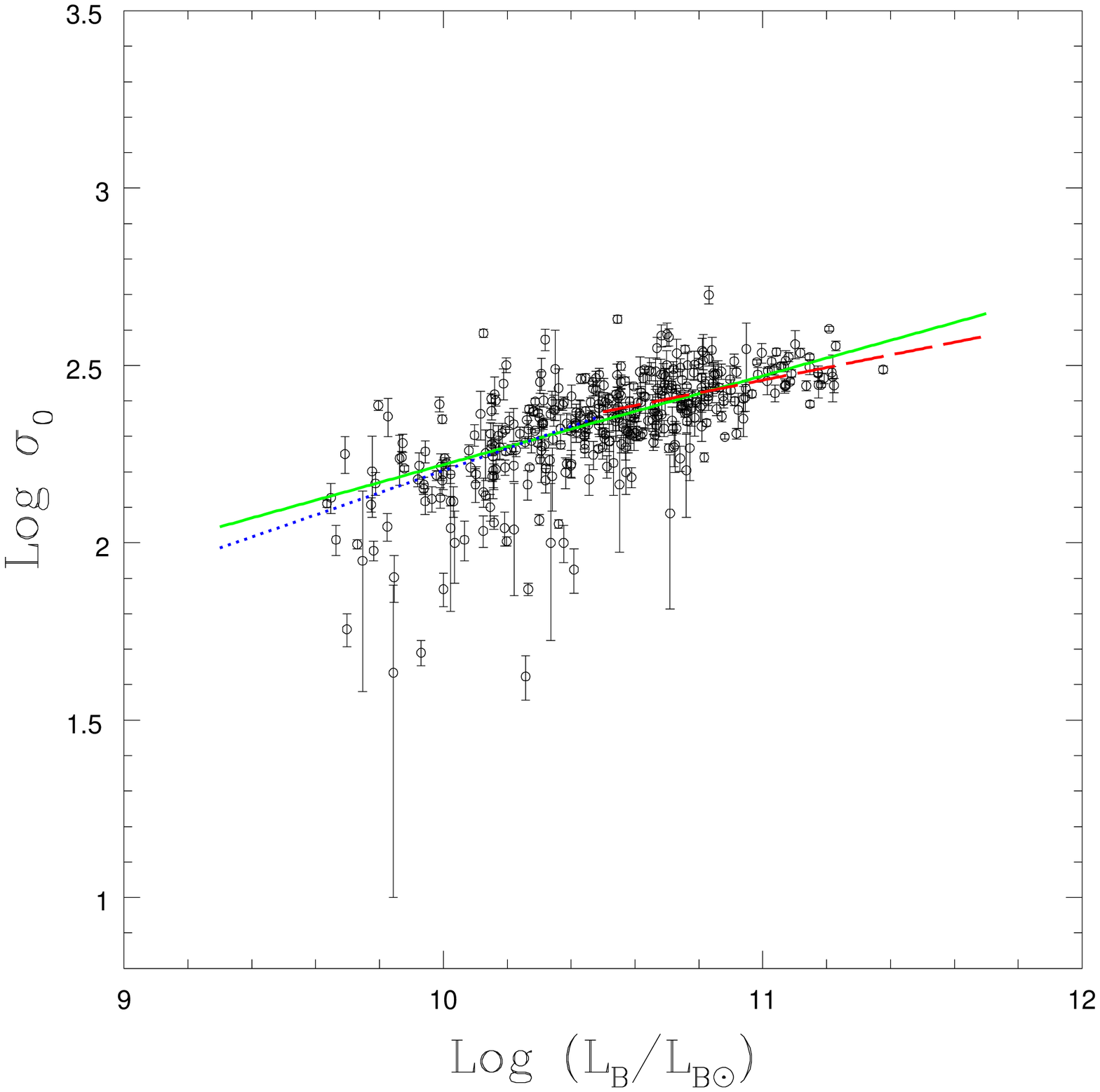]{The FJ relation for the whole sample. The green continuos line is the best fit for the whole sample ($L_{\rm B}$ $\propto$
 $\sigma_0^{4.0}$). The red dashed ( $L_{\rm B}$ $\propto$ $\sigma_0^{5.6}$)  and the blue dotted ($L_{\rm B}$ $\propto$ $\sigma_0^{3.2}$) lines are  the best fits for the  low ( Log($L_{\rm B}/L_{\rm B \odot})$ $<$ 10.5  )and high 
 (Log($L_{\rm B}/L_{\rm B\odot})$   $\ge$ 10.5) luminosity ellipticals. All fits have been derived accounting for errors on $\sigma_0$. \label{fig3}}  
 
 \figcaption[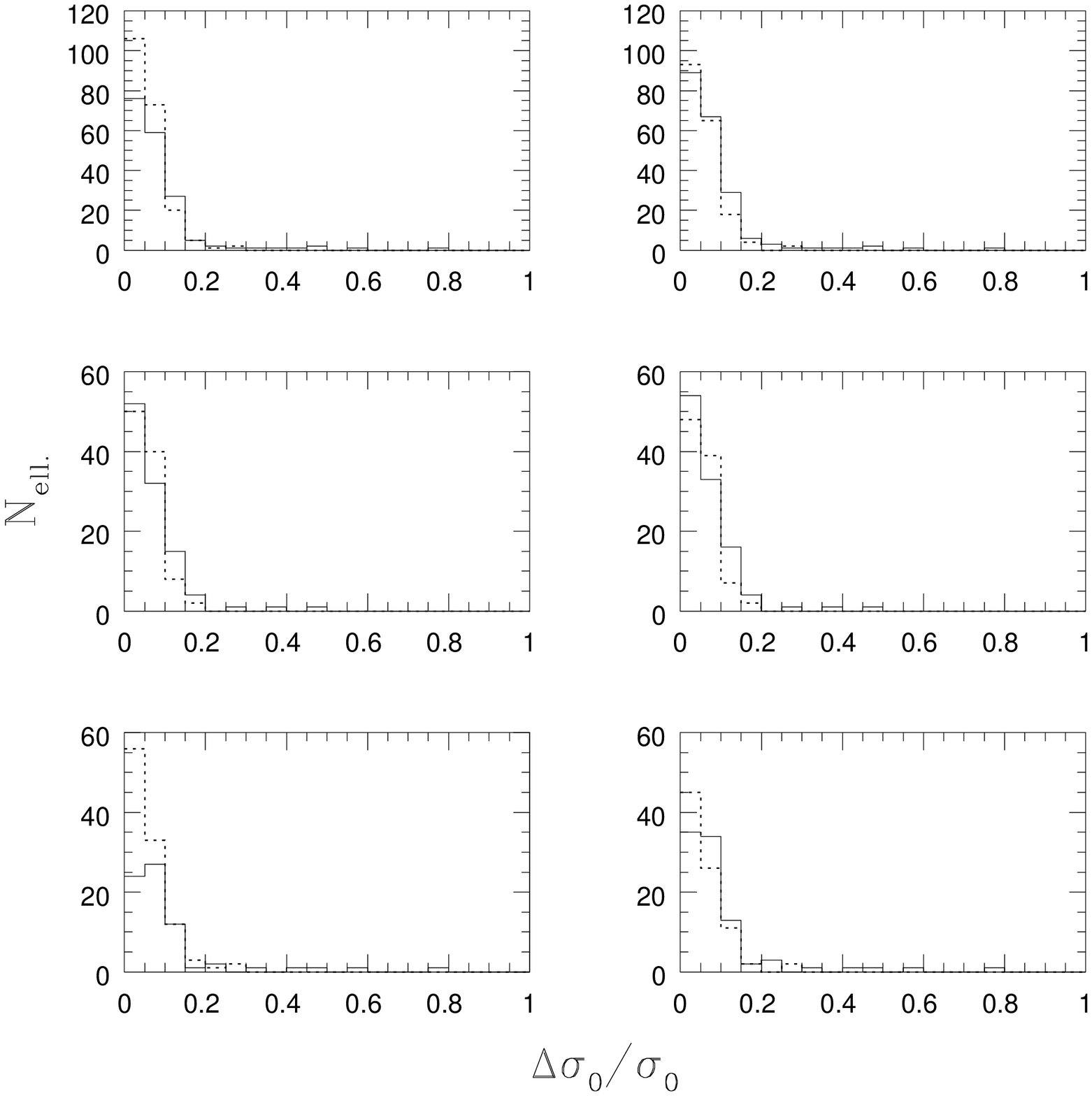]{The distribution of relative errors on $\sigma_0$  ($\Delta\sigma_0/\sigma_0$) for ellipticals laying above (dotted)
 and below (continous) the best fit line in the whole sample (upper panels) and in the high and low luminosity subsamples (middle and lower
 panels). Left panels refer to unweighted fits, right panels to fits weighted for errors on $\sigma_0$. \label{fig4}}
 
\figcaption[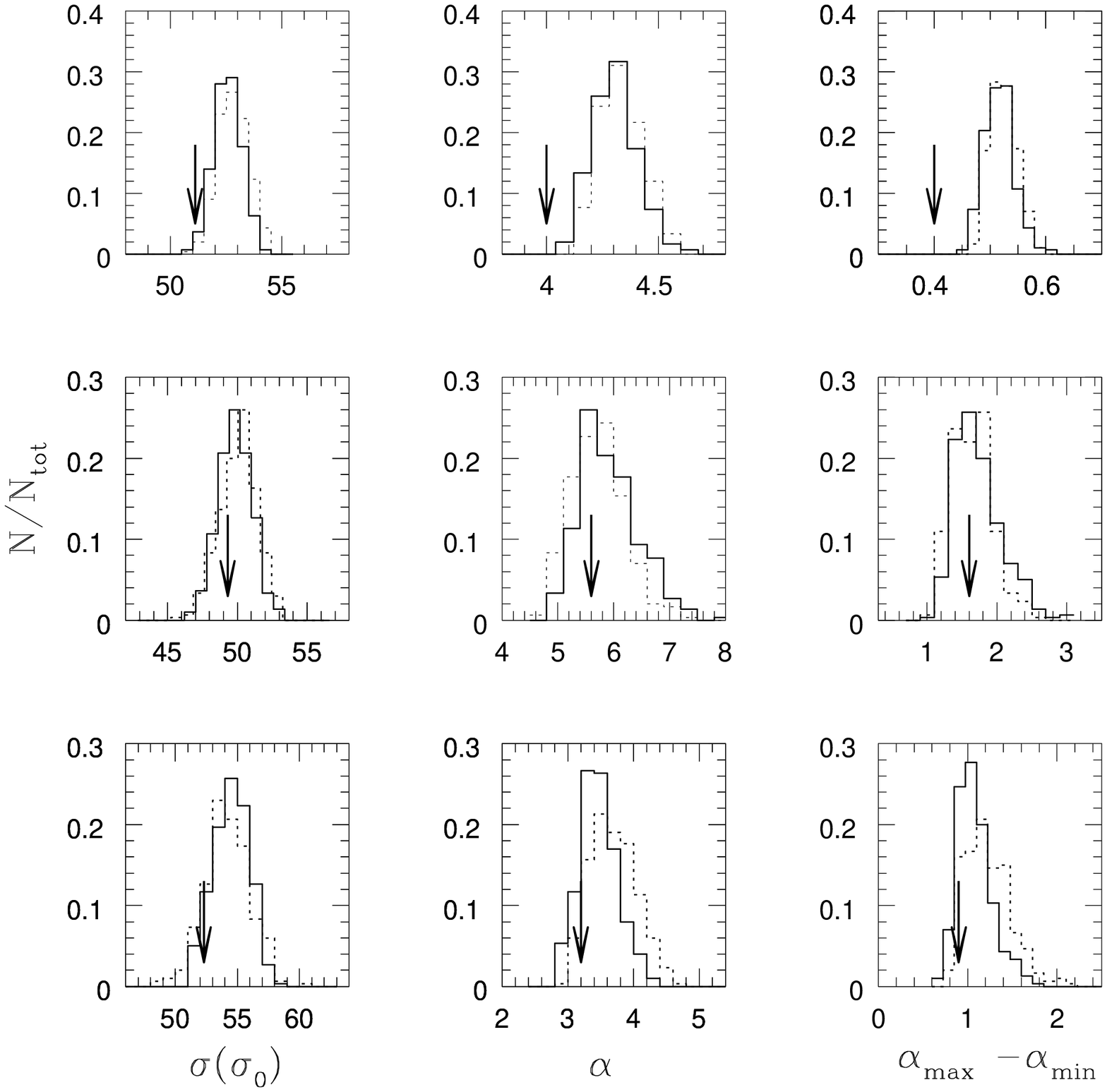]{The effect of the error on $m_{\rm B}$ on the FJ relation.  Upper panels refer to the whole sample, middle and lower panels to the  high and low luminosity
subsamples. Each distribution (dotted or continuous histograms) refers and is normalized to a set of  300 simulated samples that we have obtained randomly adding or subtracting to 
each $m_{\rm B}$ its true  (dotted histogram) or average (continuos histogram) error. The arrow on each panel indicates 
corresponding values obtained for the real sample and subsamples.
 \label{fig5}}  
 
\figcaption[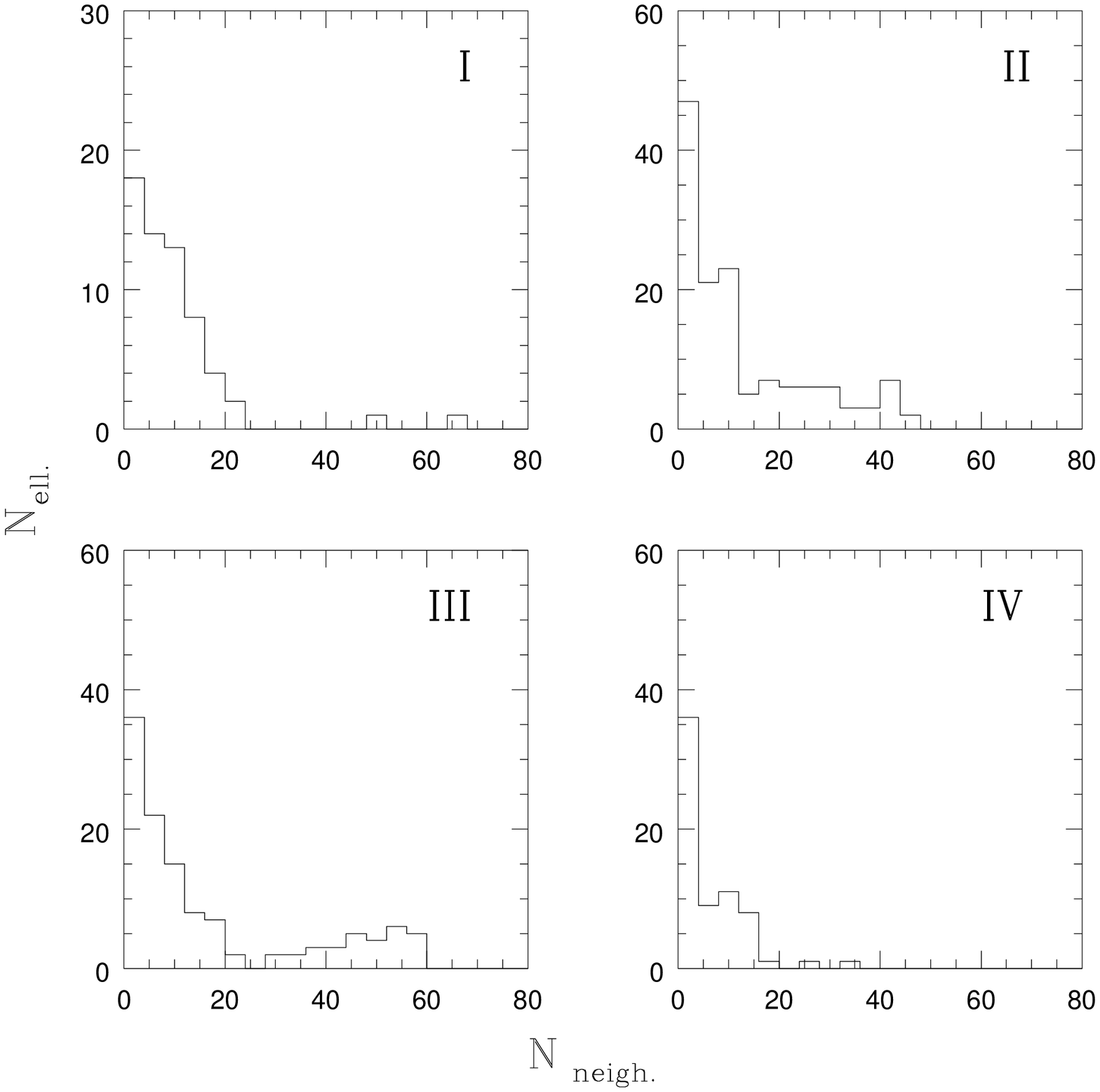]{The number of neighbours
 ($N_{\rm neigh}$) distribution  for ellipticals in each of the 4 distance bins in which we have divided the whole
 sample (cfr. Table 2).\label{fig6}}  
   
\figcaption[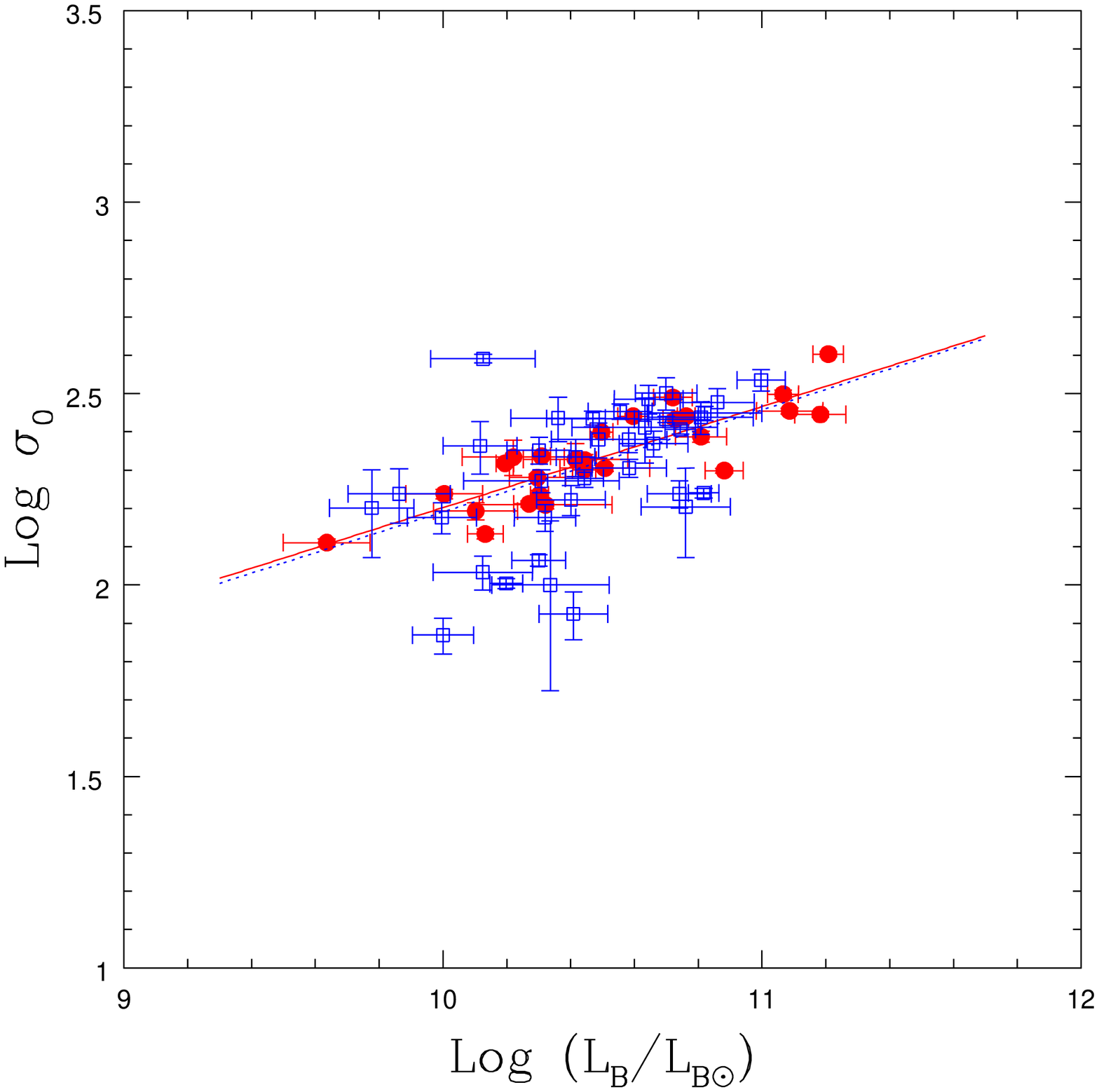]{The FJ relation for ellipticals in high density (red filled circles) and 
 low density (blue open squares) environments. Overimposed are the corresponding best fit lines (red continous
 $L_{\rm B}$ $\propto$ $\sigma_0^{3.79}$ and blue dotted $L_{\rm B}$ $\propto$ $\sigma_0^{3.75}$).\label{fig7}}
 
\figcaption[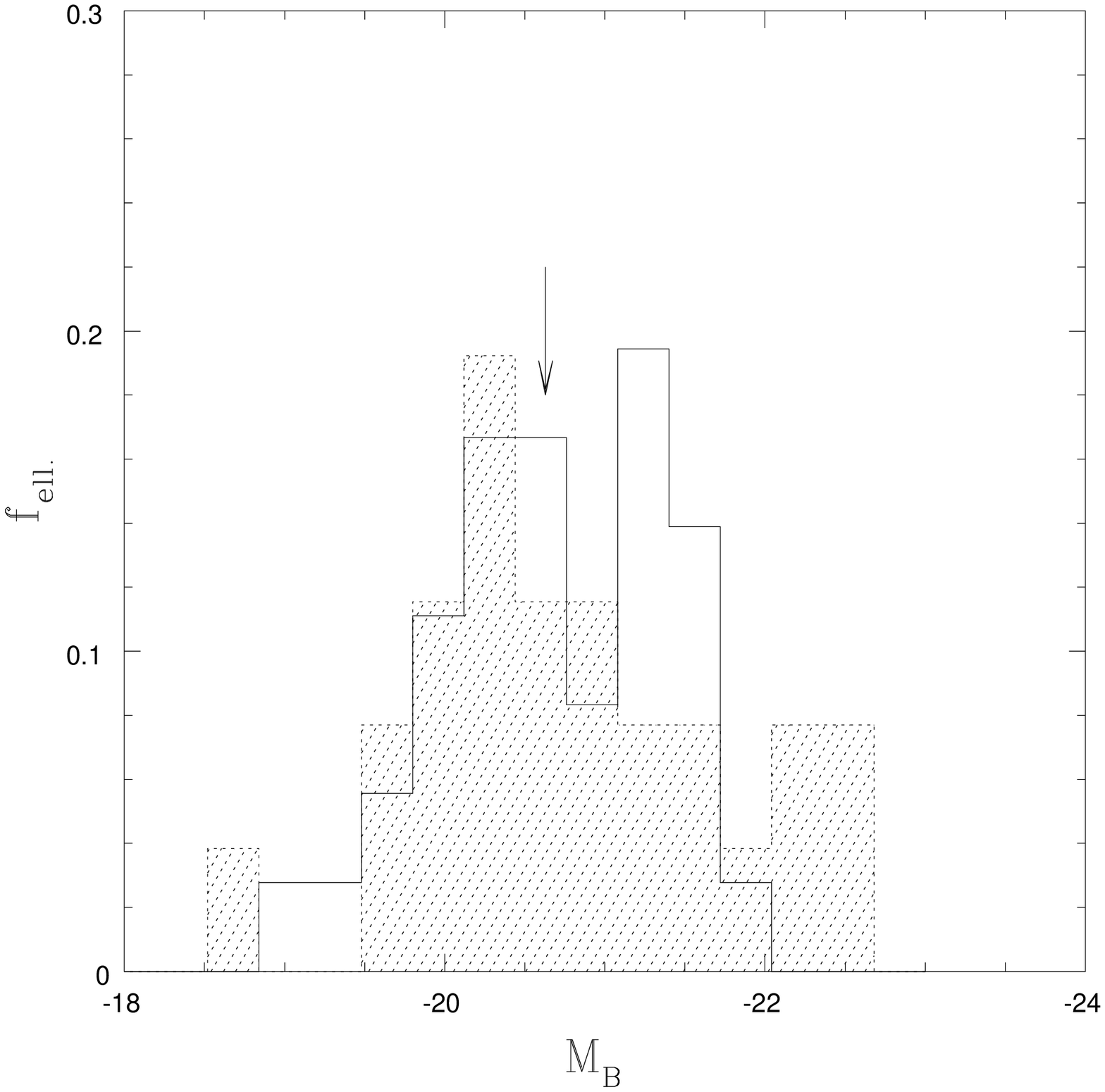]{$M_{\rm B}$ distribution of ellipticals in low and high  
  (shaded) density environments. Both distributions are normalized to the total number of ellipticals
  in each sample. The arrow indicates the median value of $M_{\rm B}$ which is the same for both samples. \label{fig8}}
  
 \figcaption[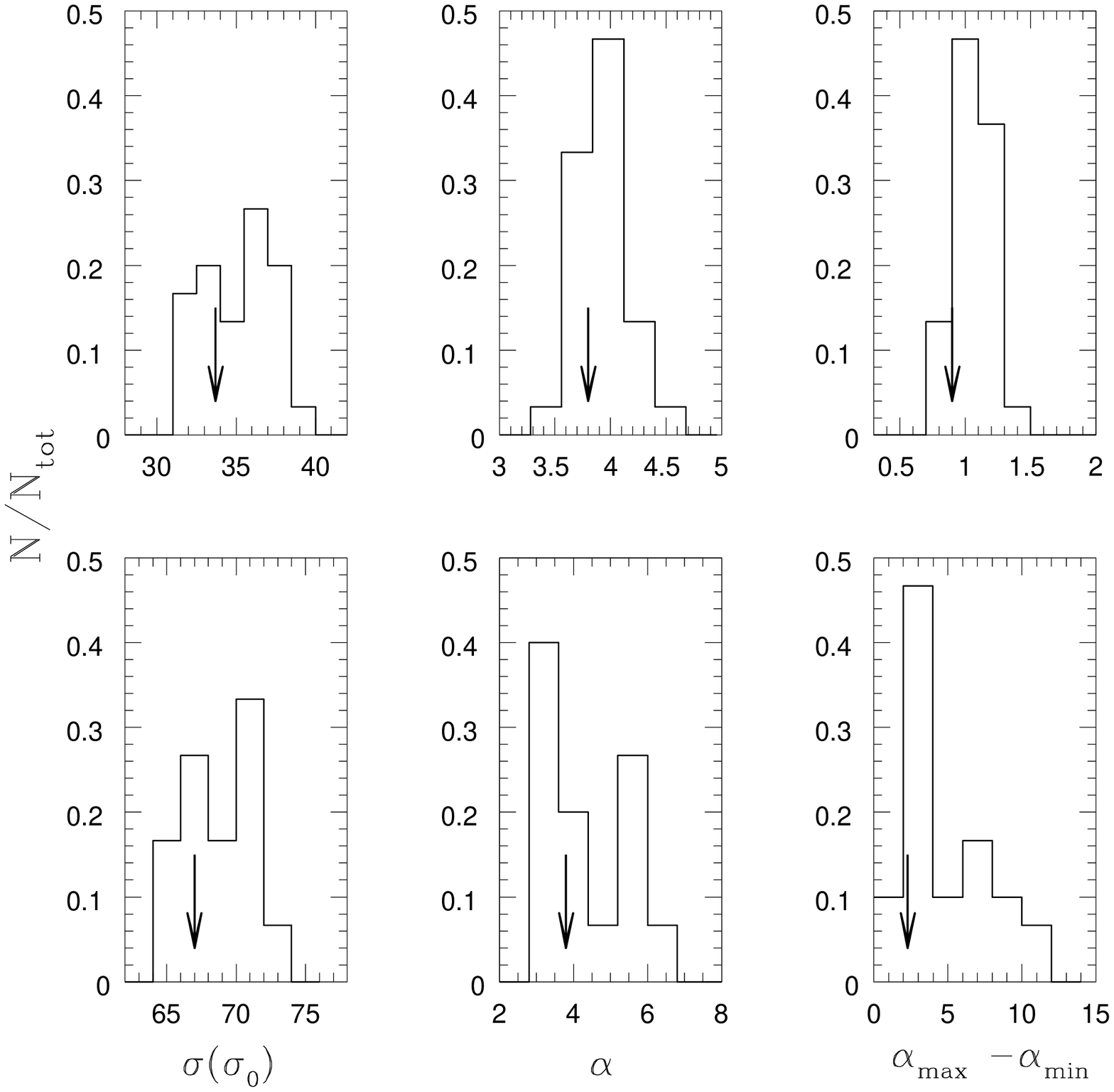]{The effect of the error on $m_{\rm B}$ on the FJ relation for the high density (upper panels) and low density (lower panels) environment.    Each distribution  refers and is normalized to a set of  30 simulated samples that we have obtained randomly adding or subtracting to 
each $m_{\rm B}$ its true error. The arrow on each panel indicates corresponding values obtained for the real samples. \label{fig9}}

\clearpage

\plotone{fig1.eps}

\clearpage

\plotone{fig2.eps}

\clearpage
 
 \plotone{fig3.eps}
 
 \clearpage
 
 \plotone{fig4.eps}
 
 \clearpage
 
 \plotone{fig5.eps}
 
 \clearpage
 
 \plotone{fig6.eps}
 
 \clearpage

 \plotone{fig7.eps}
 
 \clearpage

 \plotone{fig8.eps}
  
  \clearpage
 
 \plotone{fig9.eps}
 
 
 
 
\end{document}